\documentclass[apjl]{emulateapj_arxiv}

\usepackage{graphicx}
\usepackage{color}

\shortauthors{Karak \& Nandy}
\shorttitle{Turbulent Pumping and Solar Cycle Memory}

\begin{document}
\title{Turbulent Pumping of Magnetic Flux Reduces Solar Cycle Memory and thus Impacts Predictability of the Sun's Activity}
\author{Bidya Binay Karak}
\affil{Department of Physics, Indian Institute of Science, Bangalore 560012, India}
  \email{bidya\_karak@physics.iisc.ernet.in}
\and
\author{Dibyendu Nandy}
\affil{Indian Institute for Science Education and Research, Kolkata, Mohampur 741252, West Bengal, India}
  \email{dnandi@iiserkol.ac.in}

\begin{abstract}
Prediction of the Sun's magnetic activity is important because of its effect on space environment and climate. However, recent efforts to predict the amplitude of the solar cycle have resulted in diverging forecasts with no consensus. Yeates et al. (2008) have shown that the dynamical memory of the solar dynamo mechanism governs predictability and this memory is different for advection- and diffusion-dominated solar convection zones. By utilizing stochastically forced, kinematic dynamo simulations, we demonstrate that the inclusion of downward turbulent pumping of magnetic flux reduces the memory of both advection- and diffusion-dominated solar dynamos to only one cycle; stronger pumping degrades this memory further. Thus, our results reconcile the diverging dynamo-model-based forecasts for the amplitude of solar cycle 24. We conclude that reliable predictions for the maximum of solar activity can be made only at the preceding minimum--allowing about 5 years of advance planning for space weather. For more accurate predictions, sequential data assimilation would be necessary in forecasting models to account for the Sun's short memory.

\end{abstract}

\keywords{magnetic fields -- Sun: activity -- Sun: dynamo -- Sun: interior}
\newcommand{\Bf}{{\bf B}}
\newcommand{\vf}{{\bf v}}
\newcommand{\vp}{{\bf {v_p}}}
\newcommand{\pa}{\partial}
\newcommand{\Rs}{R_{\odot}}
\newcommand{\er}{\mbox{erf}}
\newcommand{\Bc}{B_c}
\def\Rs{R_{\odot}}

\section{Introduction}
Sunspots are strongly magnetized regions on the solar surface and have been systematically observed over the last four centuries. These observations show that the magnetic output of the Sun is roughly periodic, with the number of sunspots on the solar surface increasing and decreasing in a cyclic fashion with a mean period of 11 years. Sunspots are the seats of solar magnetic storms--solar flares and coronal mass ejections--that eject large amounts of magnetized plasma into space generating severe space weather. When these storms are Earth directed, they pose a serious hazard to satellites, air-traffic on polar routes and modern-day technologies such as telecommunication, GPS networks and electric power grids at high latitudes (US National Academy of Sciences 2008). The slow, long-term evolution of magnetic activity also affects the radiative output of the Sun (Lean 1997; Krivova et al. 2007)--the primary natural energy input into the climate system, and is therefore important from the perspective of climate dynamics. This realization has driven many recent efforts to predict the activity of the Sun, however, with disappointingly, diverging results (Pesnell 2008).

The magnetic activity of the Sun is generated through a magnetohydrodynamic (MHD) dynamo mechanism involving interactions between plasma flows and magnetic fields within the solar convection zone (SCZ). The current understanding (Charbonneau 2010; Nandy 2012) of the solar cycle is that the strong toroidal ($\phi$) component of the solar magnetic field is generated in the solar interior through stretching of the poloidal ($r$-$\theta$) component of the magnetic field by the differentially rotating solar plasma. These toroidal flux tubes are magnetically buoyant and can rise from the interior and erupt through the solar surface forming bipolar sunspot pairs. Traditionally two mechanisms--namely the mean-field $\alpha$-effect driven by helical turbulence, and the Babcock--Leighton process (Babcock 1961; Leighton 1969; 
Mu\~noz-Jaramillo et al. 2010) related to the decay and dispersal of tilted bipolar sunspot pairs at near-surface layers--are thought to recreate the poloidal component back from the toroidal field. Here, we focus on the Babcock--Leighton solar dynamo model as the Babcock--Leighton process can be constrained by surface magnetic field observations and long-term observations spanning multiple solar cycles suggests that the Babcock--Leighton mechanism predominantly contributes to the solar poloidal field (Dasi-Espuig et al. 2010).

Since the two components of the Sun's magnetic field are produced at spatially segregated layers in the SCZ, transport of magnetic flux between them is necessary for the functioning of the cycle.  The finite time required for magnetic flux transport introduces a memory into the system which allows for dynamo model based solar cycle predictions. While magnetic buoyancy plays the dominant role in transporting toroidal field to the surface, there are diverse competing mechanisms for transporting the poloidal field generated at near-surface layers back to the solar interior. In advection-dominated models meridional circulation is assumed to be the primary mechanism, while in diffusion-dominated models turbulent diffusion is assumed to be the primary mechanism for achieving the transport of poloidal field into the generating layer for the toroidal component. Amongst the many predictions (Pesnell 2008) 
till date for the amplitude of solar cycle 24, the only two physics (dynamo) based forecasts disagree strongly. One predicts a very strong cycle (Dikpati et al. 2006) while the other a very weak cycle (Choudhuri {\it et al.} 2007). Yeates {\it et al.} (2008) have shown that the diverging predictions result from different assumption by these predictive models regarding the nature of the SCZ; while Dikpati {\it et al.} (2006) assume a advection-dominated SCZ, Choudhuri {\it et al.} (2007) assume a diffusion-dominated SCZ. Yeates {\it et al.} (2008) have demonstrated that the memory of the solar dynamo is longer in the advection-dominated case (lasting over several solar cycles) as opposed to the diffusion-dominated case (lasting over one cycle) and this they argue is the origin of the disagreement in the predictions (see also Jiang {\it et al.} 2007). Solar observations cannot conclusively determine whether the Sun's convection zone is diffusion-dominated or advection-dominated, further compounding the problem of solar cycle predictions.

However, another important flux transport mechanism in the SCZ, namely the turbulent 
pumping of magnetic flux, has not received adequate attention and has been completely 
ignored in solar cycle prediction models. Theoretical analysis and magneto-convection 
simulations by several authors 
(Petrovay \& Szakaly 1993; Brandenburg et al. 1996; Tobias et al. 1998, 2001; 
Dorch \& Nordlund 2001; Ossendrijver et al. 2002;
Ziegler \& R\"udiger 2003; K\"apyl\"a et al. 2006a; Rogachevskii et al. 2011; 
Pipin \& Seehafer 2009; Seehafer \& Pipin 2009) 
suggest that magnetic fields in the SCZ are pumped preferentially along radial direction with maximum pumping speeds of a few m/s. This radial pumping is particularly effective on the weak fields which are not subject to strong buoyancy forces, mainly acts vertically downward and presumably results from a combination of factors including the presence of strong, coherent downward plumes and radial gradient in turbulence. An equatorward directed latitudinal pumping, mainly concentrated near low latitudes, also arises when the rotation becomes important (Ossendrijver et al. 2002; K\"apyl\"a et al. 2006a). 
Many studies indicate that turbulent pumping plays a dynamically important role in the context of the solar and the stellar dynamo mechanism
(Brandenburg et al. 1992; K\"apyl\"a et al. 2006b; Guerrero \& de Gouveia Dal Pino 2008;
Kitchatinov \& R\"udiger 2012; Do Cao \& Brun 2012; Cameron et al. 2012). 
Nevertheless, the effect of turbulent pumping on solar cycle memory and predictability has been hitherto unexplored.

\section{Model}
Here, within the domain of a stochastically forced kinematic solar dynamo model, in addition to meridional circulation and turbulent diffusion mediated transport of the poloidal field, we explore the effects of vertically downward turbulent pumping of poloidal magnetic field as a dominant flux transport mechanism. The evolution of the poloidal and toroidal components of the solar magnetic fields in the framework of the ($\alpha$-$\Omega$) kinematic solar dynamo model are given by\\

\begin{equation}
\frac{\pa A}{\pa t} + \frac{1}{s}[(\vf + \vp).\nabla][s A]
= \eta_{p} \left( \nabla^2 - \frac{1}{s^2} \right) A + \alpha B
\label{pol_eq}
\end{equation}
\\
\begin{eqnarray}
\frac{\pa B}{\pa t}
+ \frac{1}{r} \left[ \frac{\pa}{\pa r}
(r v_r B) + \frac{\pa}{\pa \theta}(v_{\theta} B) \right]
= \eta_{t} \left( \nabla^2 - \frac{1}{s^2} \right) B \nonumber \\
+ s(\Bf_p.\nabla)\Omega + \frac{1}{r}\frac{d\eta_t}{dr}
\frac{\partial}{\partial{r}}(r B)~~~~~
\end{eqnarray}
with $s = r \sin \theta$. Here $A$ is the vector potential for the poloidal component of the magnetic field ($\Bf_p $=$ \nabla \times A$) and $B$ is the toroidal magnetic field. The internal rotation of the Sun denoted by $\Omega$ is determined from an analytic fit to helioseismic data. Meridional circulation is given by $\vf = v_r {\bf{\hat{e}}_r} + v_{\theta} {\bf{\hat{e}}_{\theta}}$; it is directed poleward near the surface layers, equatorward near the base of the SCZ and is subject to mass conservation (i.e., satisfies ${\nabla}.{\rho}{\vf} = 0$). The poloidal field source-term $\alpha$--which is concentrated at near-surface layers--models the Babcock--Leighton mechanism of poloidal field creation. Effective magnetic diffusivities for the poloidal and toroidal component of the field are denoted by $\eta_p$ and $\eta_t$, respectively. We solve the above dynamo equations with appropriate choices of parameters and boundary conditions within a $r$-$\theta$ domain that includes the SCZ and one solar hemisphere. The typical set-up of this model has been well-studied (Nandy \& Choudhuri 2002; Chatterjee {\it et al.} 2004; Yeates {\it et al.} 2008) and for ease of comparison--specifically in the context of solar cycle memory, here we use the same parameters as utilized in the study of Yeates {\it et al.} (2008)--where all input parameters have been described in details.

Turbulent pumping appears as an advective term in the mean field induction equation, is shown to be more effective on the weaker poloidal fields, and previous studies indicate that it is directed downwards across all latitudes 
(Ossendrijver et al. 2002; K\"apyl\"a et al. 2006a). 
The stronger toroidal component of the field which exists as localized and magnetically buoyant flux tubes in the SCZ is expected to be affected less by the pumping mechanism. The memory of the solar cycle--exploring which is the primary aim of this study--is also known to be affected primarily by the time delay introduced due to the transport of poloidal flux downward into the solar interior. Motivated by these considerations, and to keep our simulations and the interpretation of the results tractable, we include only the effects of downward turbulent pumping on the poloidal component of the magnetic field. To achieve this we add an extra term in the advective part of Eq.~1, given by ${\vp} = {\gamma_r}{{\bf{\hat{e}_r}}}$, to model radial pumping. Note that this parameterization in not intended to model a global downward movement of plasma, but only that of magnetic flux and hence does not need to satisfy mass conservation. Guided by numerical simulations (Ossendrijver et al. 2002; K\"apyl\"a et al. 2006a) we use the following profile for $\gamma_r$:
\begin{eqnarray}
\gamma_r = - f \gamma_{0r} \left[ 1 + \rm{erf}\left( \frac{r - 0.715}{0.015}\right) \right] \left[ 1 - \rm{erf} \left( \frac{r-0.97}{0.1}\right) \right] \nonumber \\
\times \left[ \rm{exp}\left( \frac{r-0.715}{0.25}\right) ^2 \rm{cos}\theta +1\right] ~~~~
\end{eqnarray}
The value of $f \gamma_{0r}$ determines the amplitude of $\gamma_r$, which we vary in our simulations to explore its effects on solar cycle memory. The radial pumping profile in the SCZ based on Eq.~3 is depicted in Fig.~1. It is negative (i.e., vertically downward) throughout the convection zone and vanishes below $0.7R$ (near the base of the SCZ). Its value is maximum near the poles at a radius around $0.9R$ and decreases towards the equator and in the solar interior.

\begin{figure}[!h]
\epsscale{1.0}
\plotone{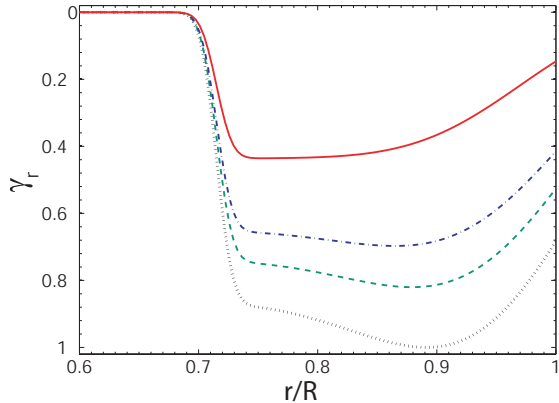}
\caption{Dotted, dashed, dot-dashed and solid lines show the variation of radial pumping $\gamma_r$ at co-latitudes $\theta = 0^0$, $45^0$, $60^0$ and $90^0$, respectively, with fractional solar radius from the solar surface to the solar interior.}
\end{figure}

\section{Simulations}
Fluctuations in the solar cycle in the context of the Babcock--Leighton dynamo arise from the random buffeting of rising magnetic flux tubes, resulting in a scatter around the mean tilt angle of bipolar sunspot pairs given by Joy's law--which is in fact observed (Dasi-Espuig et al. 2010). Thus, to imbibe the stochastic nature of the Babcock--Leighton poloidal field generation process in our model, we introduce random fluctuations in the $\alpha$-effect by setting $\alpha_0 = \alpha_{\rm{base}} + \alpha_{\rm{fluc}} \sigma (t, \tau_{\rm{cor}})$, with $\alpha_{\rm{base}}$ = 30 m s$^{-1}$. For our simulations we take $\alpha_{\rm{fluc}} = \alpha_{\rm{base}}$ (i.e., with fluctuations up to $100\%$). Here, $\sigma$ is an uniform random deviate between [$-1$ and $+1$]. The coherence time $\tau_{cor}$ is chosen such that there are around ten fluctuations in a (sunspot) cycle. Treating stochastic fluctuations in this manner is standard practice, even in dynamo models based on the mean-field $\alpha$-effect models (Pipin {\it et al.} 2012). Although a precise observational constraint on the amplitude of the Babcock--Leighton $\alpha$-effect does not exist, an order of magnitude estimate is possible based on the efficacy of the decay and dispersal of active region flux governed by diffusion and meridional circulation. For the diffusive dispersal velocity ($v_\eta = \eta / L$), considering an active region scale $L = 10^7$ m and $\eta = 1\times10^{8}$ m$^2$~s$^{-1}$, yields $v_\eta = 10$ m s$^{-1}$. In conjunction, a typical near-surface meridional circulation speed on the order of $20$ m s$^{-1}$ motivates our choice of the Babcock--Leighton $\alpha$-effect amplitude.

With stochastic forcing, we explore the effect of turbulent pumping on solar cycle memory through simulations in two different regimes, diffusion-dominated ($v_0 = 15$ m~s$^{-1}$, $\eta_0 = 1\times10^{12}$ cm$^2$~s$^{-1}$) and advection-dominated regime ($v_0 = 26$ m~s$^{-1}$, $\eta_0 = 1\times 10^{12}$ cm$^2$~s$^{-1}$). With these parameter choices, in the former case turbulent diffusivity plays a dominant role, while in the latter case advection by meridional circulation plays a dominant role in the transport of poloidal flux from the upper to the lower layers of 
the SCZ (Yeates et al. 2008). We vary the maximum pumping speed
progressively from 0 to 4 m s$^{-1}$ in both these regimes, 
allowing transport by turbulent pumping to become increasingly important in the dynamics of magnetic flux transport.

\section{Results}

\begin{figure}
\epsscale{1.2}
\plotone{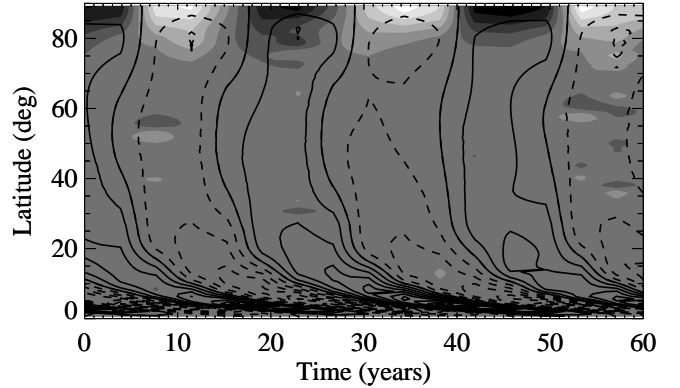}
\caption{Representative butterfly diagram of the solar toroidal field at the base of the convection zone (solid and dashed contours depict positive and negative fields, respectively) is overlaid on the radial field at the solar surface
(in grayscale, with white denoting positive and black denoting negative values). This diagram depicts the spatio-temporal field evolution from a stochastically forced dynamo simulation in the advection-dominated regime with a pumping speed of 2 m~s$^{-1}$. These simulations exhibit cyclic reversals as well as fluctuations in the magnetic field amplitude.}
\label{bfly_ap2}
\end{figure}

Figure~\ref{bfly_ap2} shows a butterfly diagram of the toroidal field at the base of the convection zone overlaid on the surface radial field from a simulation in the advection-dominated regime, with a pumping speed of 2 m~s$^{-1}$. The amplitude of the generated toroidal and poloidal fields vary from one cycle to another in these stochastically forced runs. Over continuous simulations spanning multiple solar cycles we calculate the polar radial flux $\phi_{r}$ located within the latitude $70^0$ to $89^0$ at the solar surface and the toroidal flux $\phi_{tor}$ located within a layer $r=0.677 R_{\odot}$--$0.726 R_{\odot}$ and latitudinal extent $10^0$--$45^0$ -- i.e., in the low latitude tachocline. From this we calculate the peak value of $\phi_{tor}$ and $\phi_{r}$ of each cycle to compute the correlation between the peak surface radial flux for cycle $n$ with the toroidal flux of subsequent cycles. The extent of this correlation, as argued by Yeates {\it et al.} (2008), establishes the dynamical memory of the solar cycle. Note that this approach does not have to rely on identifying the precise epoch of solar maximum or minimum based on any qualitative appraisal of the butterfly diagram and is therefore both quantitative and objective. The simulated toroidal flux for a cycle is on the order of $10^{25}$ Mx which is close to what is observed but the simulated peak polar flux of about $10^{24}$ Mx is higher than the observed polar flux at solar minimum of about $10^{22}$ Mx. However, the latter observations are based on relatively low-resolution magnetograms and the recent high resolution observations of unipolar kilogauss flux tubes in the polar region (Tsuneta {\it et al.} 2008) can potentially reconcile this (for detailed arguments see Choudhuri 2003 and the online supplementary information in Nandy et al. 2011).

The influence of increasingly dominant turbulent pumping on the memory of the solar cycle is summarized in Table~1 for both advection- and diffusion-dominated convection zones. Spearman's rank correlation coefficients and significance levels have been computed between the peak polar and toroidal fluxes of various cycles. As a consistency check, for the case without any turbulent pumping, we first recover the result of Yeates {\it et al.} (2008) that the memory of the solar dynamo is short in the diffusion-dominated regime--with the correlation lasting only between the polar flux of cycle (n) and the toroidal flux of cycle (n+1), but long in the advection-dominated regime--where the poloidal flux of cycle (n) correlates with the toroidal flux of cycles (n), (n+1), (n+2) and (n+3) with the memory degrading slowly.

\begin{table}[t]
\caption[]{Correlation coefficients ($r_s$) and percentage significance levels ($p$) for peak surface radial flux $\Phi_{\rm{r}}$ of cycle $n$ versus peak toroidal flux $\Phi_{\rm{tor}}$ of different cycles for 275 solar cycles data. The first column denotes the amplitude of the turbulent pumping speed in various simulation studies. The top row corresponds to the case without turbulent pumping and subsequent rows corresponds to simulations with increasing pumping speeds.}
  \begin{center}\begin{tabular}{lrrrrr}
 \hline
                               &              &Dif. Dom.                                       & Adv. Dom.      \\
 \hline
Pumping                        & Parameters~~~~~~~                                 & $r_s$ ($p$)& $r_s$~($p$) \\
\hline
                               & $\Phi_{\rm{r}}(n)~\&~\Phi_{\rm{tor}}(n)$~~~~~~& $0.19~(99.9)$ & $0.57~(99.9)$\\
                               & $\Phi_{\rm{r}}(n)~\&~\Phi_{\rm{tor}}(n+1)$    & $0.64~(99.9)$ & $0.77~(99.9)$\\ [-1ex]
\raisebox{1.0ex}{0 m s$^{-1}$} & $\Phi_{\rm{r}}(n)~\&~\Phi_{\rm{tor}}(n+2)$    & $0.04~(55.9)$ & $0.46~(99.9)$\\
                               & $\Phi_{\rm{r}}(n)~\&~\Phi_{\rm{tor}}(n+3)$    & $0.22~(99.9)$ & $0.27~(99.9)$\\
\hline
                               & $\Phi_{\rm{r}}(n)~\&~\Phi_{\rm{tor}}(n)$~~~~~~& $-0.06~(67.0)$ & $0.41~(99.9)$\\
                               & $\Phi_{\rm{r}}(n)~\&~\Phi_{\rm{tor}}(n+1)$    & $0.67~(99.9)$ & $0.72~(99.9)$\\[-1ex]
\raisebox{1.0ex}{1 m s$^{-1}$} & $\Phi_{\rm{r}}(n)~\&~\Phi_{\rm{tor}}(n+2)$    & $0.09~(83.9)$ & $0.29~(99.9)$\\
                               & $\Phi_{\rm{r}}(n)~\&~\Phi_{\rm{tor}}(n+3)$    & $-0.02~(26.5)$ & $-0.01~(18.9)$\\
\hline
                               & $\Phi_{\rm{r}}(n)~\&~\Phi_{\rm{tor}}(n)$~~~~~~& $0.12~(94.9)$ & $0.19~(99.8)$\\
                               & $\Phi_{\rm{r}}(n)~\&~\Phi_{\rm{tor}}(n+1)$    & $0.43~(99.9)$ & $0.75~(99.9)$\\[-1ex]
\raisebox{1.0ex}{2 m s$^{-1}$} & $\Phi_{\rm{r}}(n)~\&~\Phi_{\rm{tor}}(n+2)$    & $-0.16~(99.9)$ & $0.07~(73.8)$\\
                               & $\Phi_{\rm{r}}(n)~\&~\Phi_{\rm{tor}}(n+3)$    & $-0.02~(20.8)$ & $-0.10~(89.8)$\\
\hline
                               & $\Phi_{\rm{r}}(n)~\&~\Phi_{\rm{tor}}(n)$~~~~~~& $0.11~(49.2)$ & $0.29~(92.0)$\\
                               & $\Phi_{\rm{r}}(n)~\&~\Phi_{\rm{tor}}(n+1)$    & $0.32~(99.9)$ & $0.62~(99.9)$\\[-1ex]
\raisebox{1.0ex}{3 m s$^{-1}$} & $\Phi_{\rm{r}}(n)~\&~\Phi_{\rm{tor}}(n+2)$    & $-0.18~(99.6)$ & $0.07~(78.0)$\\
                               & $\Phi_{\rm{r}}(n)~\&~\Phi_{\rm{tor}}(n+3)$    & $0.03~(36.6)$ & $-0.10~(91.6)$\\
\hline
                               & $\Phi_{\rm{r}}(n)~\&~\Phi_{\rm{tor}}(n)$~~~~~~& $0.19~(99.8)$ & $0.30~(99.9)$\\
                               & $\Phi_{\rm{r}}(n)~\&~\Phi_{\rm{tor}}(n+1)$    & $0.26~(99.9)$ & $0.46~(99.9)$\\[-1ex]
\raisebox{1.0ex}{4 m s$^{-1}$} & $\Phi_{\rm{r}}(n)~\&~\Phi_{\rm{tor}}(n+2)$    & $-0.16~(99.3)$ & $0.07~(72.8)$\\
                               & $\Phi_{\rm{r}}(n)~\&~\Phi_{\rm{tor}}(n+3)$    & $-0.10~(91.9)$ & $-0.22~(99.9)$\\
\hline
\end{tabular}
  \end{center}
\end{table}

\begin{figure*}
\epsscale{0.6}
\plotone{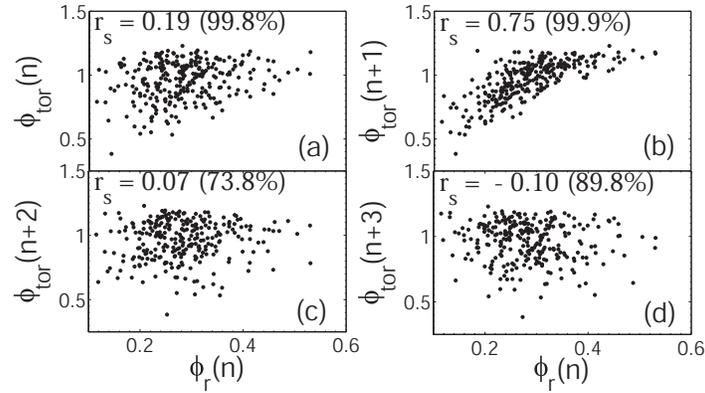}
%\vspace{-1.0cm}
\caption{Relationship between the peak (polar) radial flux $\phi_{\rm{r}}(n)$ and the peak toroidal flux $\phi_{\rm{tor}}$ of cycle (a) $n$ (b) $n+1$, (c) $n+2$, and (d) $n+3$ in the advection-dominated regime with a pumping speed amplitude of 2 m~s$^{-1}$. The flux values are in units of $10^{25}$ Mx. The Spearman's rank correlation coefficients ($r_s$) along with significance levels are inscribed. The dynamical memory persists only between the peak polar flux of cycle (n) and the peak toroidal flux of cycle (n+1), i.e., spanning a time which is less than one solar cycle.}
\label{corr_ap2}
\end{figure*}

\begin{figure*}
\epsscale{0.6}
\plotone{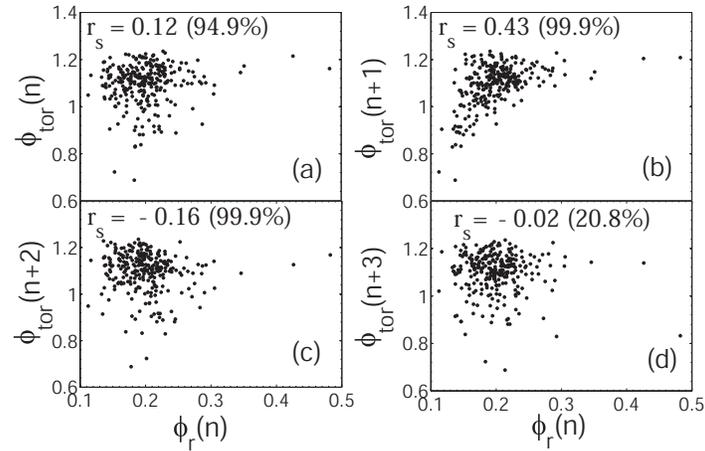}
%\vspace{-1.0cm}
\caption{Same as Fig.~3, but for the diffusion-dominated regime.}
\label{corr_dp2}
\end{figure*}

However, it is observed that when turbulent pumping becomes increasingly important, the advection-dominated dynamo system starts losing its memory and at a modest pumping speed of 2 m~s$^{-1}$, the significant correlation is only between the polar flux of cycle (n) and the toroidal flux of the next cycle (n+1). Unexpectedly therefore, the diffusion-dominated and advection-dominated dynamo regimes become similar--at least in terms of memory--when flux transport by turbulent pumping becomes significant. Figure~\ref{corr_ap2} (for the advection-dominated regime) and Fig.~\ref{corr_dp2} (for the diffusion-dominated regime), where we plot the relationship between poloidal and toroidal flux of various cycles with a pumping speed of 2 m s$^{-1}$, reinforce this conclusion. Moreover, as summarized in Table~1, we find that when the amplitude of turbulent flux pumping is made stronger the memory of the solar cycle reduces for all cases, and for a pumping speed of 4 m s$^{-1}$ even the one cycle memory is severely degraded. To check the parameter dependence of the results we have performed some simulations with changes in meridional circulation in conjunction with fluctuations in the poloidal field generation process. Preliminary findings indicate that the above results remain qualitatively unchanged.

\section{Implications for Solar Cycle Predictions}

The overreaching conclusion from this study is that the memory of the solar cycle is short, extending over a time period barely spanning one solar cycle (between the minimum in activity of one cycle and the maximum in activity of the next). Our analysis, which leads to that conclusion is remarkable in that it resolves the ambiguity that existed between solar cycle predictions based on the differing memory of advection- and diffusion-dominated solar dynamos. Expressed succinctly, the relative efficiency of advective or diffusive flux transport, by meridional circulation or turbulent diffusion, respectively, becomes less significant in the presence of turbulent pumping of magnetic flux and the latter now governs the dynamical memory of the system. This is because turbulent pumping dominates the vertical flux transport by coupling the two spatially segregated source layers of the solar dynamo.

A discussion on the timescales corresponding to competing flux transport processes in the SCZ illuminates the physics underlying our results. A meridional flow speed of $v_0 = 26~$m~s$^{-1}$ results in a advective flux transport timescale ($L/v_0$, with $L$ estimated along a typical stream-line looping through the pole) ranging from 10 to 20 yr to transport the surface poloidal field to the mid-to-low latitude tachocline--where the toroidal field spawns sunspots of the next cycle. Turbulent diffusion with $\eta_0 = 1\times10^{12}~$cm$^2~$s$^{-1}$ results in a diffusion timescale ($L^2 /\eta_0$, where $L$ signifies the depth of the SCZ) of 14 yr. On the other hand, even a modest turbulent pumping velocity of $\gamma_r = 2$~m~s$^{-1}$ results in a comparably much shorter, pumping timescale ($L/\gamma_r$, where $L$ signifies the depth of the SCZ) of about 3.4 yr. Thus, the dynamical memory of the solar cycle becomes very short on inclusion of turbulent pumping. Faster pumping results in an even shorter memory, which becomes comparable to the coherence time of stochastic fluctuations in the near-surface poloidal field generation process. This results in fluctuations of the poloidal field, say between the minimum of cycle (n) and the maximum of cycle (n+1), being reflected in the tachocline where the toroidal field of cycle (n+1) is stored and amplified, thereby, degrading even the one cycle memory of the system.

In summary, we find that the solar cycle memory is short, extending only over one cycle; the poloidal field at the minimum of a given cycle may be used to predict the amplitude of the next sunspot cycle. This allows for space weather preparedness and planning about 5 years in advance, but presumably no longer--at least based on this current understanding. Nonetheless, even one cycle forecasts may be fraught with difficulties as efficient turbulent flux pumping from the surface to the solar interior can degrade this one cycle memory. Perhaps, in retrospect, this explains why multiple attempts--even the precursor-based ones--to forecast the amplitude of solar cycle 24 have diverged. In the light of these findings, we believe that the best strategy for accurate solar cycle forecasts lies in data assimilation techniques that can compensate for the loss of memory through data assimilation at frequent intervals in forecasting models, with the requirement that those intervals be relatively shorter than the memory of the solar cycle.

\begin{acknowledgements}
This research was supported by the Ramanujan Fellowship of the Department of Science and Technology, the Ministry of Human Resource Development, the Council for Scientific and Industrial Research of the Government of India and the NASA Living With a Star Program. BBK thanks IISER Kolkata for hosting him during the performance of this work.
\end{acknowledgements}

\def\apj{{ ApJ,}}
\def\aa{{ A\&A,}}
\def\sol{{ Sol.\ Phys.,}}
\def\mnr{{Mon.\ Not.\ R.\ Astron.\ Soc.,}}

\end{document}